\documentclass[usenatbib]{mn2e}
\usepackage{epsfig}
\usepackage{hyperref}
\hypersetup{
 colorlinks,
citecolor = [rgb]{0,0.5,0} }
\usepackage{amsmath}
\def\gtwid{\mathrel{\raise.3ex\hbox{$>$\kern-.75em\lower1ex\hbox{$\sim$}}}}
\def\ltwid{\mathrel{\raise.3ex\hbox{$<$\kern-.75em\lower1ex\hbox{$\sim$}}}}
\def\\{\hfil\break}
\def\ie{{\it i.e. }}
\def\eg{{\it e.g. }}

\def\lesssim{\mathrel{\hbox{\rlap{\hbox{\lower4pt\hbox{$\sim$}}}\hbox{$<$}}}}
\def\gtrsim{\mathrel{\hbox{\rlap{\hbox{\lower4pt\hbox{$\sim$}}}\hbox{$>$}}}}

\newcommand{\mamo}[1]{\mbox{$#1$}}
\newcommand{\unit}[1]{\ifmmode \:\mbox{\rm #1}\else \mbox{#1}\fi}
\newcommand{\mone}{\mamo{^{-1}}}

\newcommand{\kms}{\unit{km~s\mone}}

\newcommand{\mpc}{\unit{Mpc}}

\newcommand{\hmpc}{\mamo{h\mone}\mpc}

\begin{document}

\title[Interpretable Bulk Flows]{Easily Interpretable Bulk Flows: Continuing Tension with the Standard Cosmological Model}
\vskip 0.5cm
\author[Peery, Watkins,\& Feldman]{Sarah Peery$^{\dagger}$, Richard Watkins$^{\dagger,1}$ \& Hume A. Feldman$^{\star,2}$\\
$^\dagger$Department of Physics, Willamette University, Salem, OR 97301, USA.\\
$^\star$Department of Physics \& Astronomy, University of Kansas, Lawrence, KS 66045, USA.\\
emails: $^1$rwatkins@willamette.edu;\, $^2$feldman@ku.edu}

\maketitle
\begin{abstract}
We present an improved Minimal Variance (MV) method for using a radial peculiar velocity sample to estimate the average of the three-dimensional velocity field over a spherical volume, which leads to an easily interpretable bulk flow measurement.   The only assumption required for this interpretation is that the velocity field is irrotational.   The resulting bulk flow estimate is particularly insensitive to smaller scale flows.   We also introduce a new constraint into the MV method that ensures that bulk flow estimates are independent of the value of the Hubble constant $H_o$; this is important given the tension between the locally measured $H_o$ and that obtained from the cosmic background radiation observations.    We apply our method to the \textit{CosmicFlows-3} catalogue and find that, while the bulk flows for shallower spheres are consistent with the standard cosmological model, there is some tension between the bulk flow in a spherical volume with radius $150$\hmpc\ and its expectations; we find only a $\sim 2\%$ chance of obtaining a bulk flow as large or larger in the standard cosmological model with \textit{Planck} parameters.
\end{abstract}

\begin{keywords}
Cosmology: Cosmological parameters; large-scale structure of Universe; Theory; Galaxies: distances and redshift; kinematics and dynamics; statistics
\end{keywords}

\section{Introduction}
\label{sec:intro}

One of the central ideas of the standard model of cosmology is the gravitational instability paradigm, in which small amplitude density fluctuations in the early Universe were amplified  
by gravity into the large-scale structure that we see today \citep{MukChi81}.   In this picture, growth of fluctuations continues today, taking the form of flows of galaxies on scales of 1 Mpc and larger.   On scales of order 100 Mpc, these flows are still in the linear regime, and thus can be directly related to density fluctuations on these scales.   Large-scale flows are thus an important cosmological probe that can potentially provide confirmation of many features of the standard model.  

Given that measurements of peculiar velocities of individual galaxies have large uncertainties, researchers have traditionally focused on velocity field statistics that average over many galaxies; by far the most common statistic is the dipole moment of the velocity field, also called the bulk flow.     Early attempts to measure the bulk flow
 \citep[\eg][]{RubThoForRob76,DreFabBurDav87,LP94,RiePreKir95} were plagued by small sample sizes and difficult to control biases.  Recently, however, the compilation of large catalogues of distance measurements with $\sim 10,000$ objects   \citep[\eg][]{CF3} has made it possible to accurately estimate the bulk flow on scales $\sim 100$\hmpc\ \citep{ WatFelHud09,FelWatHud10,DavNusMas11,NusDav11,Nusser14,WatFel15,ScrDavBlaSta15,FeiBraNus17,HelNusFeiBil17,HelBilLib18}. For recent discussions of bulk flow determination methods, see \eg, \citet{DavScr14,WatFel15a,Nusser16}. 
 
Bulk flows have additional significance as a large-scale flow statistic in being, at least in principle, independent of the value of the Hubble constant, $H_o$.   While calculated peculiar velocities depend strongly on the Hubble constant, since bulk flows measure the dipole of the velocity field, they are theoretically insensitive to phantom monopole flows introduced by using an incorrect value of $H_o$.    This characteristic of bulk flows is particularly appealing in an era where there is tension between cosmic background radiation measurements of $H_o$ and more local measurements using the Hubble diagram (see, \eg, \citet{RieMarHofSco16,BerVerRie16,PlanckXLVIII16,Freedman17,RieCasYuaMar18}).

This paper is organized as follows: in Section~\ref{stats} we discuss some of the different ways of defining the bulk flow and the difficulty of comparing bulk flow estimates reported in different studies.     In Section \ref{sec:theory} we follow \citet{Nusser14} and show how the assumption of an irrotational flow allows the full, three-dimensional bulk flow to be written in terms of a weighted average of  the radial component of the peculiar velocity, the only component of the velocity that is actually observable.  In Section~\ref{sec:MV}, we show that the resulting weighted average is well suited for estimation via the Minimal Variance (MV) method.  We also show how a new constraint can be added to the MV method to ensure that the bulk flow estimate is  independent of the value of the Hubble constant.  Section \ref{sec:data} discusses the data on which we apply our method. In Section \ref{sec:results} we present the results of our analysis.   Finally, in section \ref{sec:discussion}, we discuss our results and put them in the context of other estimates of the bulk flow.  

\section{Bulk Flow Estimation}
\label{stats}

While the bulk flow is easy to understand as a concept, measurements of the bulk flow can be difficult to interpret, and are usually not comparable between studies.   How the calculated bulk flow of a survey probes motions on different scales is dependent not just on the size of a survey, but in how the individual sample velocities are weighted in the analysis.   For example, peculiar velocity measurements typically have uncertainties that grow rapidly with distance.   Given that nearby galaxies are also overrepresented in catalogues due to the relative ease of their observation, we have much more information about local flows than those at the outer regions of our samples.   Unless this imbalance of information is accounted for, measured bulk flows can end up reflecting much smaller scales than that of the survey being used.  One way to quantify precisely how the bulk flow probes the power spectrum is to calculate its window function as we  discuss below \citep[See also][]{MacFelFer11,MacFelFerJaf12}.    
 
Given the long history of bulk flow measurement, it is perhaps surprising how much confusion surrounds its definition, estimation and interpretation.  When we think of the bulk flow, we typically envision the average of the full, three-dimensional peculiar velocity $v_i$ over a spherical volume $V$ with a radius $R$
\begin{equation}
U_i = \frac{1}{V}\int_V v_i\ d^3 r,
\label{eq:bf}
\end{equation}
where $i=x,y,z$ are the cartesian components of the velocity field.  

While this definition is quite simple in principle, in practice measuring the bulk flow defined in this way is difficult.  First, the velocity field is not measured directly, but rather through the motions of individual galaxies, used as ``tracers" of the velocity field.   The distribution of the galaxies in peculiar velocity surveys can be very nonuniform, due to both the actual distribution of galaxies and the selection function of the survey, making it very difficult to achieve the uniform integral over the volume required by Eq.~\ref{eq:bf}.   Second, peculiar velocity measurements have large uncertainties, particularly at large distances, also making it difficult to probe a volume in a uniform way.      Finally, only the radial velocity of a galaxy can be measured, making it impossible to carry out an integral of the full three-dimensional velocity field directly.   

These difficulties have led many researchers to consider an alternative framing of the bulk flow components $U_i$ as the leading-order terms in a Taylor series expansion of the local peculiar velocity field $v_i({\bf r})$ \citep{Kai88,JafKai95,FelWat08,FelWatHud10}
\begin{equation}
 v_i({\bf r})=  U_i + U_{ij}r_j+...,
\label{eq:bf-alt}
\end{equation}
where $U_{ij}$ is the shear tensor, and, following the Einstein summation convention, repeated indices are summed over.   Unlike in the integral definition given in Eq.~\ref{eq:bf}, here the bulk flow components are parameters in a model of the velocity field, allowing them to be estimated, for example, using maximum likelihood methods.   However, since this definition doesn't reference a particular volume, estimates of the bulk flow defined in this way are difficult to interpret, and are not typically comparable between peculiar velocity surveys.   The MV method \citep{WatFelHud09,FelWatHud10,WatFel15} remedies this problem by allowing velocity measurements to be weighted in a way that references an ``ideal" survey with a given geometry, thus standardizing bulk flow estimates.    In both the maximum likelihood and the MV method the bulk flow can be expressed as a weighted sum of radial peculiar velocities.

\section{Theory}
\label{sec:theory}

One of the challenges of estimating the bulk flow as defined in Eq.~\ref{eq:bf} is its expression in terms of the full three-dimensional velocity field.  Recently, \citet{Nusser14,Nusser16} showed that 
if one makes the standard assumption that the velocity field is irrotational, and hence can be expressed as the gradient of a potential field,  ${\bf v}({\bf r}) = -{\bf\nabla} \phi({\bf r})$,  this bulk flow can also be expressed in terms of the radial component of the peculiar velocity.   The key idea here is that a potential field only has one degree of freedom, so that the radial component of the velocity carries all the information of the complete vector field.  

We start by expanding the potential $\phi({\bf r})$  in terms of spherical harmonics $Y_{l,m}(\theta,\phi)$,
\begin{equation}
\phi({\bf r}) = \sum_{l,m} \phi_{l,m}(r) Y_{l,m}(\theta,\phi),
\label{eq:exp}
\end{equation}
where $\phi_{l,m}(r)$ are functions only of the radial coordinate $r$.   In considering the bulk flow we are primarily interested in the $l=1$ terms in the sum.   It is convenient to express the $l=1$ spherical harmonics in terms of the cartesian components of the radial unit vector, $\hat r_i = \hat x_i\cdot \hat r$, where
\begin{align}
\hat r_x &= \sqrt{\frac{2\pi}{3}}\left( Y_{1,-1} - Y_{1,1}\right),\\
\hat r_y &= i\sqrt{\frac{2\pi}{3}}\left( Y_{1,-1} + Y_{1,1}\right),\\
\hat r_z &= \sqrt{\frac{4\pi}{3}}Y_{1,0} .
\end{align}
Using these relations we rewrite equation Eq.~\ref{eq:exp} as
\begin{equation}
\phi({\bf r}) = \phi_o(r) + \sum_i \phi_i \hat r_i + \sum_{l>1,m} \phi_{l,m}(r) Y_{l,m}(\theta,\phi),
\label{eq:cexp}
\end{equation}
where the $\phi_i$ is used to quantify the three $l=1$ dipole components of the potential.   

Following \citet{Nusser14},  we can use the divergence theorem to write the bulk flow in terms of an integral over the surface of our spherical volume $S$, 
\begin{multline}
U_i= -\frac{1}{V}\int_V \nabla_i \phi\ d^3 r=  -\frac{1}{V}\int_V {\bf\nabla}\cdot(\hat x_i \phi)\ d^3 r \\
=-\frac{1}{V}\int_{S} \phi\ \hat x_i\cdot \hat r\ R^2d\Omega = -\frac{R^2}{V}\int_{S} \phi\ \hat r_i\ d\Omega.
\end{multline}
Now we can substitute Eq.~\ref{eq:cexp} into the integral and use the orthogonality of the spherical harmonics and the $\hat r_i$ to obtain the simple expression
\begin{equation}
U_i= -\phi_i(R)/R,
\end{equation}
since, for a spherical volume, $V= 4\pi R^3/3$.   Thus the bulk flow can be directly related to the dipole components of the scalar potential evaluated at the surface of the volume.   

\cite{Nusser14} has shown that this result can be used to express the bulk flow defined in Eq.~\ref{eq:bf} as a weighted integral of the radial component of the peculiar velocity, $s= {\bf v}\cdot \hat r$, the one component of the peculiar velocity that can be observed.  We start by noting that $s$ is given by the radial component of the gradient of $\phi$,
\begin{multline}
s({\bf r}) = \frac{\partial }{\partial r}\phi ({\bf r}) = \frac{d}{dr}\phi_o(r) + 
\sum_i\frac{d}{dr}\phi_i(r) \hat r_i \\+  \sum_{l>1,m} \frac{d}{dr}\phi_{l,m}(r) Y_{l,m}(\theta,\phi).
\label{eq:sexp}
\end{multline}
We can now imagine calculating the velocity moments $\tilde U_i$ by integrating the radial peculiar velocity $s(\bf r)$ projected onto a cartesian axis over a spherical volume with a radially dependent weighting factor $w(r)$,
\begin{multline}
\tilde U_i = \frac{1}{V}\int_V w(r) s\ \hat r_i \ d^3r  \\= -\frac{1}{R^3} \int_0^R w(r) \left(\frac{d}{dr}\phi_i(r)\right) r^2dr,
\label{eq:tilU}
\end{multline}
where we have substituted Eq.~\ref{eq:sexp} and again used the orthogonality of the angular basis functions.  
Following \cite{Nusser14}, we see that the choice $w(r) = R^2/r^2$ gives
\begin{equation}
{\tilde U}_i =  -\frac{1}{R} \int_0^R \left(\frac{d}{dr}\phi_i(r)\right) dr = -\phi_i(R)/R = U_i,
\end{equation}
where we have used the fundamental theorem of calculus.   Thus from Eq.~\ref{eq:tilU} we have
\begin{equation}
U_i =  \frac{R^2}{V}\int_V  \frac{s({\bf r})\ \hat r_i}{r^2} \ d^3r.
\label{eq:rad}
\end{equation}
This gives us the important result that the bulk flow defined in Eq.~\ref{eq:bf} in terms of the full, three-dimensional velocity field can be calculated exactly from the radial component of the peculiar velocity integrated over a spherical volume with a weighting factor proportional to  $r^{-2}$.   We note that the assumption of an irrotational flow is used only in the interpretation of the bulk flow and plays no role in our analysis; even without this assumption, the velocity moment given in Eq.~\ref{eq:rad} is a useful statistic for characterizing large-scale flows.  

\section{Estimating the Bulk Flow Using the MV Method}
\label{sec:MV}

Now that the bulk flow integral has been expressed in terms of the radial peculiar velocity field, the next step is to develop a method for estimating this integral using catalogues of peculiar velocities of individual objects.     The Minimum Variance (MV) method \citep{WatFelHud09,FelWatHud10,WatFel15} is well suited for this task, since it is designed to estimate velocity moments that would be measured by a hypothetical ``ideal" survey that probes a volume in a well determined way.  


The MV method envisions an artificial ``ideal" survey that can provide a standard reference point for actual peculiar velocity surveys.  For example, the ideal survey for measuring the bulk flow integrals in Eq. \ref{eq:rad}  consists of a large number of uniformly distributed objects within the spherical volume with positions ${\bf r}_n$ and exactly measured radial velocities $s_n$, so that the bulk flow would be given by 
\begin{equation}
U_i = \frac{R^2}{N}\sum_n \hat n_{n,i} s_n/r_n^2,
\end{equation}
where $\hat n_{n,i}$ is the $i$th component of the unit vector pointing toward the $n$th object in the catalogue and $r_n$ is the distance to the $n$th object.   Alternatively, the $r^{-2}$ weighting can be achieved through the radial distribution of the ideal survey objects, \ie by having an equal number of objects per spherical shell.  For a survey with this radial distribution, the sum weighs points at different radii equally and the bulk flow integral becomes
\begin{equation}
U_i = \frac{1}{N}\sum_n \hat n_{n,i} s_n.
\end{equation}
This second option is much preferable over the first in that one can achieve the same accuracy with a much smaller ideal survey, resulting in shorter computation times.  

The goal of the MV method is to determine weights $w_{i,n}$ for an actual catalogue of $N$ measured radial peculiar velocities $S_n$, with associated measurement uncertainties $\sigma_n$, such that the weighted sums
\begin{equation}
u_i = \sum_i^N w_{i,n}S_n,
\label{eq:up}
\end{equation}
provide the best estimate of the moments $U_i$ that would be obtained from the ideal survey \citep[for details, see][]{WatFelHud09,FelWatHud10,WatFel15}.   To find the weights, the variance $\langle \left(U_i-u_i\right)^2\rangle$ is minimized subject to some constraints, which are implemented using Lagrange multipliers.   As in previous work, we impose a normalization constraint that 
$\sum_n w_{i,n}\hat n_{n,j} = \delta_{ij}$, which ensures that a spatially constant flow is correctly estimated.  Here we also consider a new constraint that guarantees that bulk flow estimates are not affected by uncertainty in the value of the Hubble parameter $H_0$.  

A tension has recently developed between measurements of the Hubble parameter $H_0$ from cosmic microwave background analysis and those from more local studies (see, \eg, \citet{Freedman17,RieCasYuaMar18}).
A value of $H_0$ is necessary to calculate peculiar velocities, and any error in $H_0$ manifests as a phantom radial flow in the sample.   This false radial flow contributes to the bulk flow estimate if the sum over radial velocities is not done in a completely isotropic way.   We can enforce the isotropy of the bulk flow sum by adding an additional constraint to our weight calculation.   Suppose that we use a Hubble parameter that differs from the true Hubble parameter by $\delta H_0$.   This error subtracts from each peculiar velocity an amount $\delta v=\delta H_0 r_n\approx cz_n\frac{\delta H_0}{H_0}$, where we have assumed that peculiar velocities are much smaller than $c$, the speed of light.     
The contribution of the error in $H_0$ to the bulk flow $U_i$ is $\sum_n w_{i,n} cz_n \frac{\delta H_0}{H_0}$.   If we enforce the constraint that
\begin{equation}
\sum_n^N w_{i,n} cz_n = 0,
\label{eq:Hconstraint}
\end{equation}
then we can be sure that our bulk flow estimate is completely independent of any uncertainty in the value of $H_0$.

All together, then, we seek to minimize the quantity
\begin{multline}
\langle (U_i-u_i)^2\rangle +\sum_j \lambda_{ij}\left(\sum_n w_{i,n} \hat n_{n,j} \right)\\ + \beta_i\left(\sum_n w_{i,n}cz_n\right),
\end{multline}
where $\lambda_{ij}$ and $\beta_i$ are Lagrange multipliers.  
Expanding out the first term and plugging in the expression for $u_i$ from Eq.~\ref{eq:up}, we can write this expression in terms of the weights $w_{i,n}$,
\begin{multline}
\langle U_i^2\rangle -\sum_n 2w_{i,n}\langle S_nU_i\rangle + \sum_{n,m} w_{i,n}w_{i,m}\langle S_nS_m\rangle\\ +\sum_j \lambda_{ij}\left(\sum_n w_{i,n} \hat n_{n,j}\right)+ \beta_i\left(\sum_n w_{i,n}cz_n\right).
\label{eq:con}
\end{multline}
We can now find the weights that minimize this expression by taking a derivative of Eq.~\ref{eq:con} with respect to $w_{i,n}$ and setting the result equal to zero,
\begin{equation}
-2\langle S_nU_i\rangle + 2\sum_m w_{i,m}\langle S_nS_m\rangle + \sum_j \lambda_{ij} \hat n_{n,j} + \beta_i cz_n=0 .
\end{equation}
Solving for the weights gives,
\begin{equation}
w_{i,n} = \sum_m G^{-1}_{nm}\left(\langle S_mU_i\rangle - {1\over 2}\sum_j \lambda_{ij} \hat n_{m,j}-{1\over 2}\beta_i cz_m\right),
\label{eq:w}
\end{equation}
where $G_{nm}\equiv \langle S_nS_m\rangle$ is the covariance matrix of the individual measured velocities.
The values of the Lagrange multipliers can be found by plugging Eq.~\ref{eq:w} into Eq.~\ref{eq:con}  and solving the simultaneous equations for $\lambda_{ij}$ and $\beta_i$:
\begin{equation}
\lambda_{ij} = \sum_k M_{ik}^{-1}\left[ \sum_{m,n} G_{nm}^{-1}(\langle S_mU_k\rangle - D_k cz_m)\hat n_{n,j}-\delta_{kj} \right],
\end{equation}
\begin{equation}
\beta_i = 2D_i - \sum_j \lambda_{ij}L_j,
\end{equation}
where 
\begin{eqnarray}
D_i &=& {1\over B}\sum_{n,m} G^{-1}_{nm}\langle S_nU_i\rangle cz_m, \\
L_i &=& {1\over B} \sum_{n,m} G^{-1}_{nm}\hat n_{n,i} \ cz_m, \\
M_{ij} &=& {1\over 2}\sum_{n,m} G^{-1}_{nm}\left(\hat n_{n,i}-L_i cz_n\right)\hat n_{m,j}\ ,
\end{eqnarray}
and 
\begin{equation}
B = \sum_{n,m} G_{nm}^{-1}cz_ncz_m .
\end{equation}
Eq.~\ref{eq:w} allows us to calculate the MV weights for measured peculiar velocities in order to estimate moments of the velocity field that would be measured by an ideal survey.   These weights depend on the covariance matrix $G_{nm}=\langle S_nS_m\rangle$ and the correlation $\langle S_nU_i\rangle$, both of which can be calculated using linear theory given a power spectrum model.   

Since the measured peculiar velocity of the $n$th object $S_n$ includes an uncorrelated error $\delta_n$, we can write $S_n = s_n + \delta_n$, where $s_n$ is the actual peculiar velocity.   Thus 
\begin{equation}
G_{nm} =  \langle s_ns_m\rangle
+ \delta_{nm}(\sigma_*^2 + \sigma_n^2),
\end{equation}
where $\delta_{nm}$ is the Kronecker delta, $\sigma_n$ is the measurement error of the $n$th object in the catalogue, and $\sigma_*$ is the velocity noise, which accounts for small-scale motions not included in the linear model.  
Next, given that the ideal moments $U_i$ are weighted sums of the exact velocities of the ideal survey, $U_i= \sum_{n^\prime} w^\prime_{i,n^\prime} s_{n^\prime}$, where the sum is now over the objects in the ideal survey with the appropriate weights.   This leads to
\begin{equation}
\langle S_mU_p\rangle = \sum_{n^\prime} w^\prime_{pn^\prime}\langle s_ms_{n^\prime}\rangle.
\end{equation} 

Both $G_{nm}$ and $\langle S_mU_p\rangle$ depend on the correlation $\langle s_ns_m\rangle$ between radial velocities of objects at positions $\bf r_n$ and $\bf r_m$.   In terms of linear theory this is given by the integral over the density power spectrum $P(k)$ multiplied by an angle-averaged window function $f_{nm}(k)$,
\begin{equation}
 \langle s_ns_m\rangle= {H_0^2\Omega_{m}^{1.1}\over 2\pi^2}\int   dk\  P(k)f_{nm}(k),
\label{eq:Rv}
\end{equation}
where 
\begin{equation}
 f_{nm}(k) = \int {d^2{\hat k}\over 4\pi} ( {\bf \hat r}_n\cdot {\bf \hat k} )( {\bf \hat r}_m\cdot {\bf \hat k} ) 
 e^{ ik\ {\bf \hat k}\cdot ({\bf r}_n - {\bf r}_m)}.
 \label{eq:fmn}
\end{equation}
We can also calculate the covariance matrix $R_{ij}$ for the estimates of bulk flow components through
\begin{multline}
R_{ij}=\langle u_i u_j\rangle = \sum_{n,m} w_{i,n}w_{j,m} \langle s_n s_m\rangle = \\
{H_0^2\Omega_{m}^{1.1}\over 2\pi^2}\int   dk\  P(k){\cal W}^2_{ij}(k),
\label{eq:Rij}
\end{multline}
where the tensor angle-averaged window function ${\cal W}^2_{ij}$ is given by 
\begin{equation}
{\cal W}^2_{ij} =  \sum_{n,m}w_{in}w_{jm}f_{mn}(k).
\label{eq:wf}
\end{equation}
The diagonal elements of the tensor window function ${\cal W}^2_{ii}$ are extremely useful in quantifying what scales contribute to the bulk flow.   In addition, a comparison of the window function for the survey estimate $u_i$ with the window function for the ideal moment $U_i$ can indicate how well a survey can estimate a given moment.   We discuss this comparison in more detail when we present our results in section \ref{sec:results}.  For a thorough numerical study of the MV formalism see \citet{AgaFelWat12}.

For the power spectrum model $P(k)$ we use the parametrization of the $\Lambda$CDM power spectrum of \citet{EisHu98}, including the effects of baryons, with the \textit{Planck} central parameters \citep{PlanckCosPar16}, $\sigma_8= 0.8159$, $\Omega_m= 0.3089$, $\Omega_b= 0.0486$, $H_0=67.74$\kms, and $n=0.9667$.  

\begin{figure}
\centering
\includegraphics[scale=0.43]{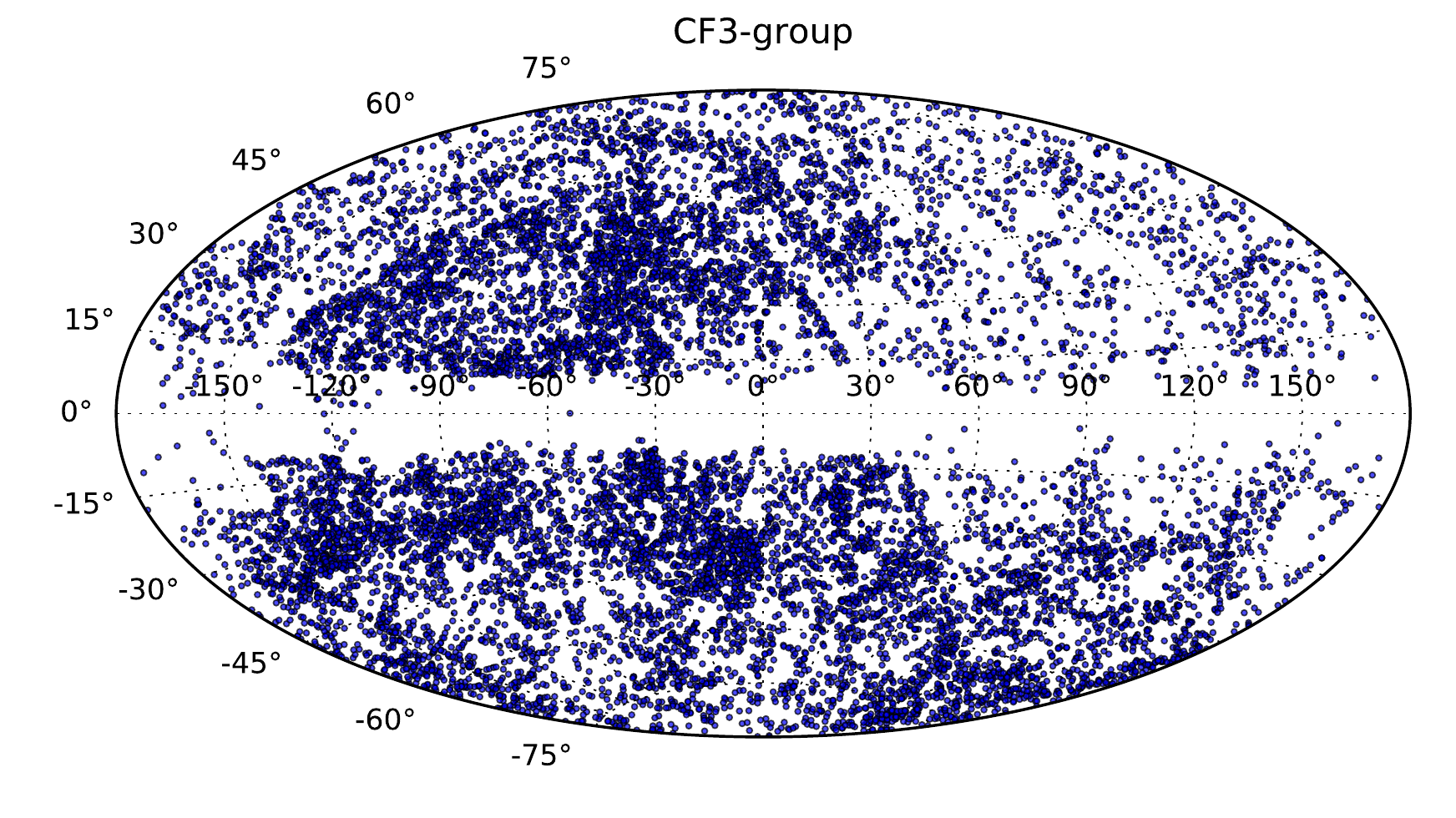}
\caption{The distribution of CF3 objects in Galactic coordinates.  }
\label{fig:hammer}
\end{figure}

\section{Data}
\label{sec:data}

In this paper we use the most recent version of the \textit{CosmicFlows} catalogue,  \textit{CosmicFlows-3} \citep{CF3}, hereafter CF3, which is a compendium of distances measured for 11,878 individual galaxies and groups, some from the literature and some from new measurements; this catalogue contains nearly all available distance measurements.     The majority of the galaxy distances are determined via the Tully-Fisher or Fundamental Plane relations, both of which give uncertainties of around 20\% of the distance, with a smaller portion of the distances coming from more accurate distance measures including SNIa, surface brightness fluctuations, Cepheids, and tip of the red giant branch.   In compiling the CF3 out of individual surveys the authors made zero-point adjustments in order to ensure consistency in the catalogue as a whole.   We use the group version of the catalogue in which galaxies determined to be in a group or a cluster have had their distances and redshifts combined into a single value of distance and redshift for the group as a whole.     

While previous versions of the \textit{CosmicFlows} catalogue \citep{TulCouDol13} were approximately isotropic, the CF3 has a markedly uneven distribution on the sky, primarily due to the addition to the catalogue of  8,885 distance measurements from the Six Degree Field Galaxy Survey (6dFGS) \citep{SprMagCol14}, all in the south celestial hemisphere.   In Fig.~\ref{fig:hammer} we show the distribution of the CF3 objects on the sky in Galactic coordinates.    The south celestial pole is roughly in the $-y$ direction in Galactic coordinates, and one can see that this side of the sky is much more heavily sampled than the $+y$ direction.   The anisotropic distribution of the CF3 catalogue objects makes it particularly important to use an analysis such as the MV method that weights objects in order to sample the volume in a well defined way;  otherwise, it would be very difficult to interpret results from a catalogue whose objects are distributed in such a nonuniform fashion.   In particular, if not properly accounted for, the anisotropic distribution of a survey can result in contributions to the bulk flow from radial flows, both real and those arising from using an incorrect value of the Hubble constant; thus demonstrating the importance of the constraint (Eq.~\ref{eq:Hconstraint}) we have introduced as discussed in Section \ref{sec:theory}.

CF3 is essentially a catalogue of distance moduli and redshifts.   Distance moduli generally have Gaussian distributed errors; however, exponentiating moduli to obtain distances skews the error distribution and leads to biased estimates of the distance $r$.  The traditional estimator of peculiar velocity, $v= cz - H_0 r$, can thus lead to biased velocities with nonGaussian error distributions.  \citet{WatFel15a} have developed a new peculiar velocity estimator $v = cz\ln (cz/H_0 r)$ that gives unbiased velocities with Gaussian errors.   Here we use this estimator to calculate peculiar velocities for the CF3 objects.   Since the MV method assumes Gaussian errors in velocities, this choice is important for ensuring the accuracy of our results.   

\begin{figure}
\centering
\includegraphics[scale=0.55]{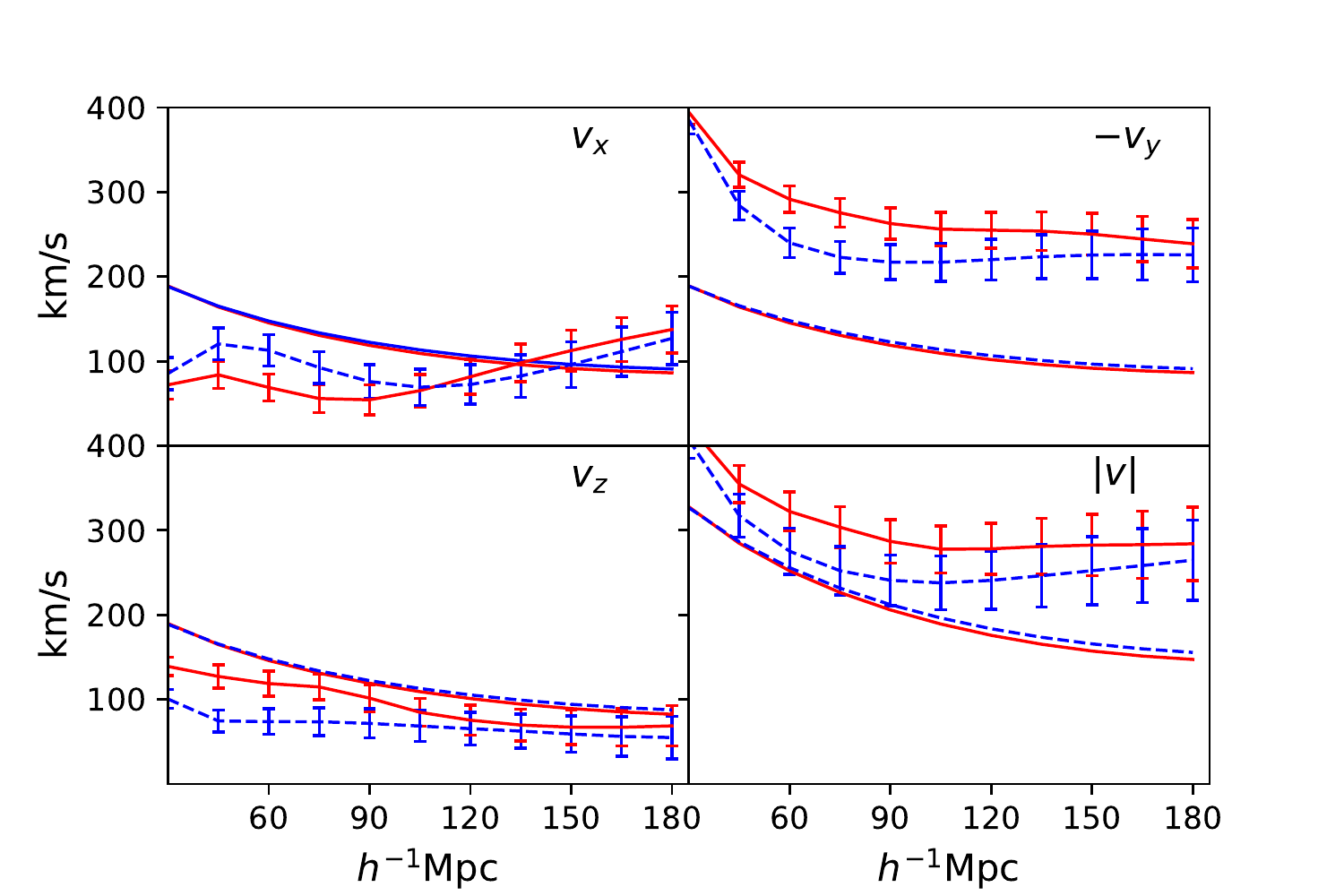}
\caption{Bulk flow components and magnitude as a function of radius $R$.  Lines with error bars are estimates from CF3-group catalogue; red solid lines are calculated using $r^{-2}$ weighting and blue dashed lines using Gaussian weighting.   The continuous lines without error bars represent expectations from the cosmological standard model using \textit{Planck} parameters.}
\label{fig:bf}
\end{figure}

In order to calculate bulk flows we also need to know the positions of the objects in the catalogue.   Although we have accurate measurements of the angular coordinates, measurements of distances are more problematic.   While it would seem natural to use the distance estimates given in the catalogue, these have large uncertainties and are prone to various types of biases.   However, the redshift $cz$ differs from $H_0 r$ only by the peculiar velocity, and provides a much more accurate measure of distance, and hence much smaller biases, for all but the closest objects in the catalogue.   Here we use $cz_n/100$\kms as the position $r_n$ of the $n$th object in units \hmpc.  This is a particularly important choice given that the CF3 catalog is redshift limited and not corrected for Malmquist bias, so that objects with large distance measurements are biased toward having negative velocities.     

\section{Results}
\label{sec:results}

In Fig.~\ref{fig:bf} we show estimates for the bulk flow components calculated using the MV method, described in Sec.~\ref{sec:MV}, with ideal surveys of varying radii $R$ weighted by $r^{-2}$.   As discussed in Sec.~\ref{sec:theory},  these estimates should correspond to the integral of the full 3-dimensional velocity field averaged over a sphere of radius $R$.   For comparison, we also show the bulk flow estimated using ideal surveys that are Gaussian balls with radial weighting 
\begin{equation}
w(r)= Ae^{-r^2/(R/3)^2},
\label{eq:gweight}
\end{equation}
where $A$ is a constant.   Note that estimates of the bulk flow components at different $R$ are provided for comparison but are not independent; since they are calculated using the same data they have highly correlated errors.  

This Gaussian weighting is similar to that used in previous MV analyses of the bulk flow \citep{WatFelHud09,FelWatHud10,AgaFelWat12,WatFel15}; however, here we have scaled the radius parameter $R$ by a factor of 3.  This is so that the resulting window function has a central peak with a similar width to the window function for $r^{-2}$ weighting with the same $R$.  We discuss this in more detail below.  

\begin{figure}
\centering
\includegraphics[scale=0.54]{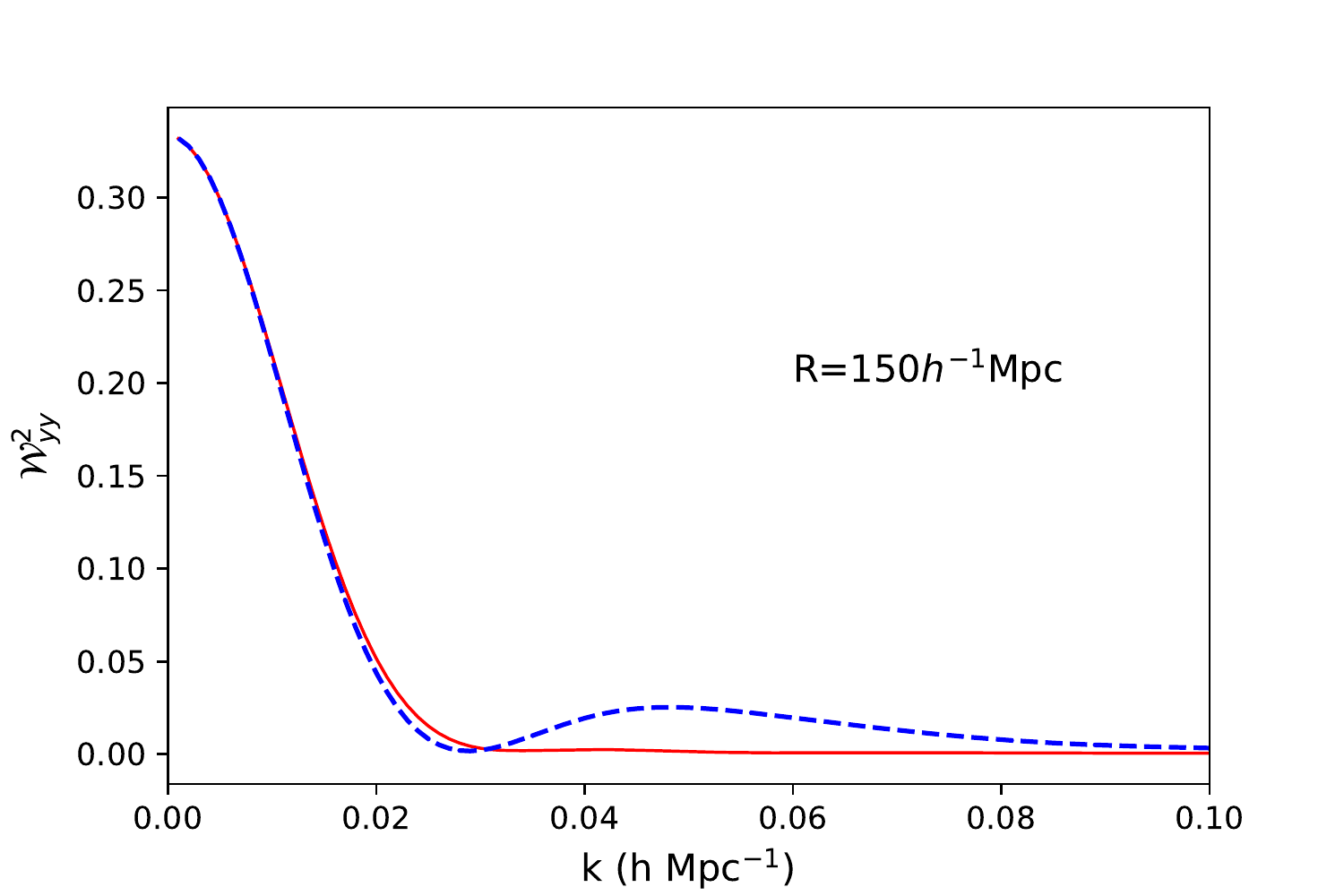}
\caption{The bulk flow window function ${\cal W}_{yy}^2$ (see Eq.~\ref{eq:wf}) from the CF3-group catalogue with $R=150$\hmpc .  The solid red line and the dashed blue line correspond to $r^{-2}$  and Gaussian weighting respectively.   Note the lack of a side-lobe in the window function using $r^{-2}$ weighting.    }
\label{fig:wincomp}
\end{figure}

In Fig.~\ref{fig:wincomp} we show window functions ${\cal W}_{yy}^2$ (the window functions for $xx$ and $zz$ are very similar) for the CF3 bulk flow estimates for the $r^{-2}$ and Gaussian weighting with $R=150$\hmpc.

The central peaks of the $r^{-2}$ and Gaussian weightings have similar widths by design; it is for this reason that we scaled the radial parameter $R$ by a factor of three in the Gaussian weighting as shown in Eq.~\ref{eq:gweight}.   This factor is necessary since the Gaussian function extends well beyond its nominal ``width" as characterized by its scale length.  Fig.~\ref{fig:wincomp} shows that the bulk flow calculated from a Gaussian ball of radius $R$ defined as in Eq.~\ref{eq:gweight} (dashed line) and a $r^{-2}$ weighted (solid line) survey in the spherical volume $r<R$  probe very similar scales.    The Gaussian weighting scheme was originally chosen to ``smooth" the sharp edge of the uniformly weighted tophat distribution and thus reduce the sidelobe of the bulk flow window function \citep{WatFelHud09} (WFH) relative to the tophat.   While it succeeds in this goal, the $r^{-2}$ weighting does even better, almost eliminating the side-lobe all together; thus bulk flows calculated using this weighting are only sensitive to velocity modes at or above the scale of the survey and are relatively insensitive to smaller scale motions within a survey.  We see that Bulk flows calculated using the $r^{-2}$ weighting scheme not only match the definition given in Eq.~\ref{eq:bf}, thus reflecting our intuition for how the bulk flow should be defined, but also have a highly desirable window function.   

\begin{figure}
\centering
\includegraphics[scale=0.54]{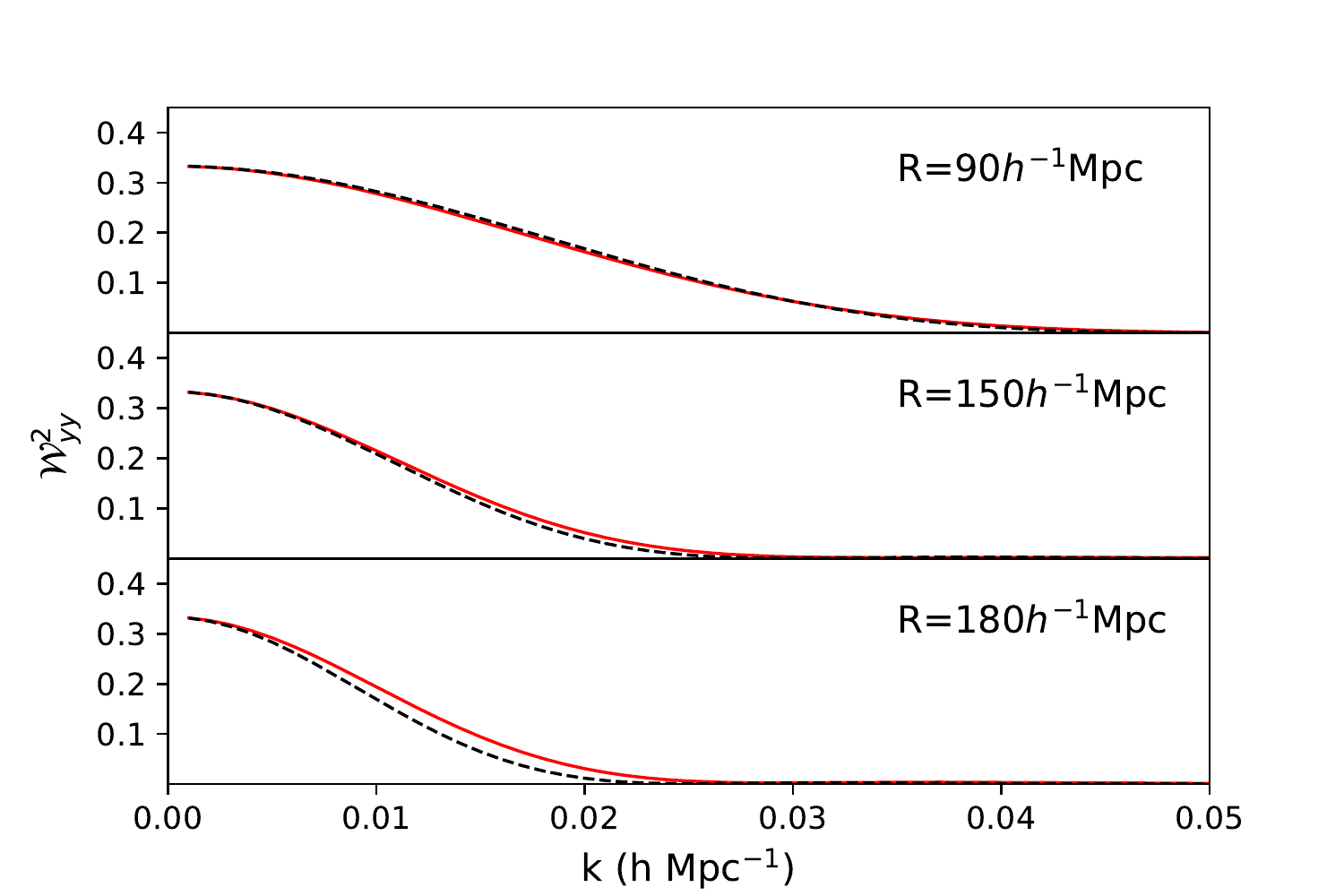}
\caption{The bulk flow window function ${\cal W}_{yy}^2$ calculated using $r^{-2}$ weighting for $R=$ 90, 150, and 180\hmpc.  The solid red line is for the CF3 catalogue and the dashed black line is for the ideal catalogue.   Note that at $R=180$\hmpc\ the CF3 window function is no longer able to match the ideal case.  }
\label{fig:winR}
\end{figure}

Fig.~\ref{fig:winR} shows the bulk  window function ${\cal W}_{yy}^2$ of the CF3 with $r^{-2}$ weighting for three different values of $R$, 90, 150, and 180 \hmpc, together with the window functions for the corresponding ideal survey.   We see from the figure that the bulk flow calculated from the CF3 using the MV method and $r^{-2}$ weighting is a good estimate of the ideal survey bulk flow for $R\lesssim 150$\hmpc,  indicating that the CF3 is well suited to estimate the bulk flow within a sphere of this radius.   However, for $R$ larger than $150$\hmpc, \eg the case of $R=180$\hmpc\ shown in the figure, we see that the central peak of the CF3 window function has reached a minimum width and has ceased to track the window function of the ideal survey.   This suggests that $R=150$\hmpc\ is the maximum radius for which the CF3 can estimate the bulk flow accurately.   

The continuous lines in Fig.~\ref{fig:bf} show the expectation for the bulk flow estimates $\sqrt{\langle u_i^2\rangle}$ calculated using the \citet{PlanckCosPar16} parameters, as discussed in Section~\ref{sec:theory}.    As would be anticipated, these expectations decrease with increasing $R$; however, the magnitude of the bulk flow doesn't decrease as fast as its expectation, so that the measured bulk flow magnitude is becoming less likely at large $R$. Beyond $R=$150\hmpc, the bulk flow remains roughly constant since there is very little additional information is being added.

\begin{figure}
\centering
\includegraphics[scale=0.54]{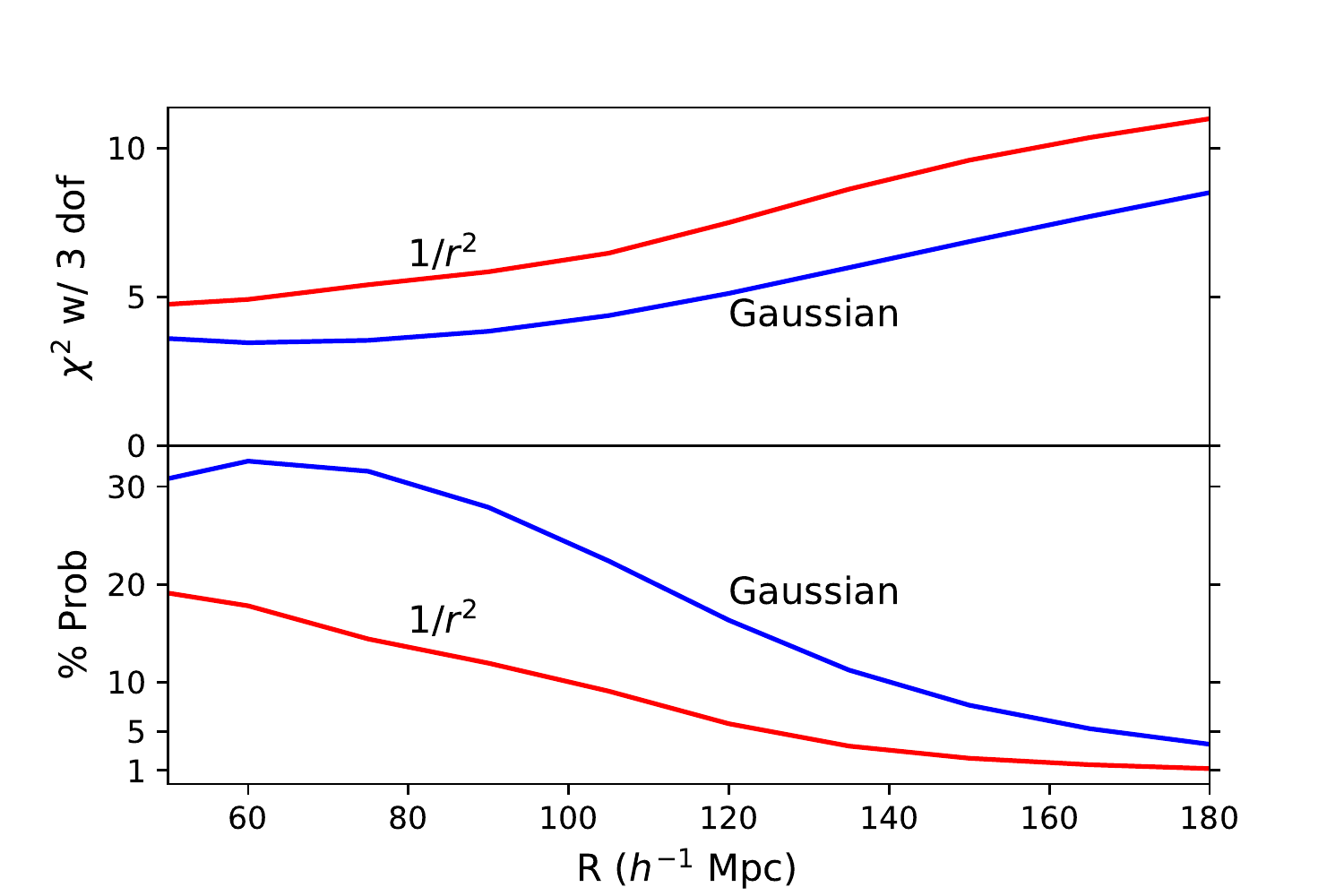}
\caption{The estimated bulk flow magnitude from CF3-group catalogue as a function of radii $R$ weighted by $r^{-2}$ }
\label{fig:chisq}
\end{figure}

A $\chi^2$ analysis can be used to quantify the probability of obtaining a bulk flow as large or larger than that measured.   The $\chi^2$ is calculated from the covariance matrix as
\begin{equation}
\chi^2 = \sum_{i,j=1}^3 u_iR_{ij}^{-1}u_j,
\end{equation}
where the covariance matrix $R_{ij}$ is given in Eq.~\ref{eq:Rij}.   In Fig.~\ref{fig:chisq} we show the $\chi^2$ calculated this way as a function of $R$, as well as the probability of obtaining a $\chi^2$ as large or larger from the $\chi^2$ distribution with 3 degrees of freedom.   

As we see from Fig.~\ref{fig:bf}, the probability for the bulk flow to be as large or larger than that measured is smallest at the largest radius we consider, $R=150$\hmpc.  We include larger $R$ in the figure for comparison, but as discussed in the text, the CF3 sample is not deep enough to accurately estimate the bulk flow on these scales.    In Table~\ref{tab:prob} we summarize the results at $R=150$\hmpc\ for both types of weightings.   

Looking at the window functions in Fig.~\ref{fig:wincomp}, it makes sense that the bulk flow expectation would be larger for the Gaussian weighting, since the Gaussian weighted bulk flows have contributions from smaller scale motions as indicated in the side lobe in its window function.    Surprisingly, though, the bulk flows calculated using Gaussian weights are consistently smaller than those using $r^{-2}$ weights.   These larger values and smaller expectations lead to the $\chi^2$ for the bulk flow using $r^{-2}$ weighting being somewhat larger than that found with Gaussian weighting.   This in turn leads to a smaller probability of finding a bulk flow as large or larger; while bulk flows calculated with Gaussian weighting are consistent with the standard model, the $r^{-2}$ weighting gives a probability of only $2.2\%$, thus indicating a tension between the bulk flow on this scale and the \textit{Planck} cosmological parameters. 

\begin{table}
\caption{Summary of Bulk Flows for $R=150$\hmpc} 
\centering
{
\begin{tabular}{lccc}
\hline
&Gaussian Wts. & $r^{-2}$ Wts. \\
Expectation (km/s) & 166 & 128\\
Bulk Flow (km/s) & 252 & 282\\
$\chi^2$ with 3 d.o.f.  &6.92 & 9.59\\
Probability &7.4 & 2.2\\
\hline
\end{tabular}
}
\label{tab:prob}
\end{table}

\section{Discussion}
\label{sec:discussion}

We have introduced a new method for calculating the bulk flow from a catalogue of peculiar velocities that has several advantages over other methods.  First and foremost, it is much easier to interpret than estimates obtained using other methods.   In particular, with the minimal assumption that the velocity field is irrotational, the bulk flow estimated with the MV method using a $r^{-2}$ weighted ideal survey corresponds to the integral of the full three dimensional velocity field over a well defined spherical volume.  

In cases where the data cannot provide an accurate estimate of the bulk flow on a particular scale, it is clearly indicated by the window function of the bulk flow estimate not matching the ideal survey window function.   In contrast, bulk flows calculated by methods such as maximum likelihood are very difficult to interpret, as they are particularly sensitive to a given survey geometry, and sample and error distributions and thus do not correspond to a well-defined volume.  Methods such as maximum likelihood give the most weight to nearby galaxies that have smaller uncertainties in their velocities, which can result in bulk flow estimates that probe smaller scales than those of the survey probes.  

A second advantage of the MV method is that it is very effective at averaging out flows on scales smaller than the volume of interest, as can be seen from the window functions in Fig.~\ref{fig:wincomp}.  Other methods of estimating the bulk flow, including other weighting schemes, can give window functions with wider central peaks and with non-negligible side lobes, resulting in bulk flow estimates that contain significant contributions from smaller scale power.    This is a very important consideration.  Bulk flow components are different from some other cosmological probes in that models do not predict their mean, which is zero, but rather their variance.   The strongest possible constraint from bulk flow measurements comes from minimizing the predicted variance; a smaller predicted variance shrinks the acceptable range of bulk flow component values for a given model.   

The primary way of reducing this variance is by increasing the depth of peculiar velocity surveys.   This has the effect of reducing the width of the central peak of the window function so that a larger fraction of bulk flow is coming from large scales, where the power spectrum vanishes.   However, a deep peculiar velocity survey does not necessarily guarantee a small predicted variance.  If a bulk flow analysis gives more weight to the inner part of the survey, where there is more information, then the central peak of the window function can be wider than expected.   Furthermore, additional variance can come from the incomplete cancellation of smaller scale flows.  The MV method with $r^{-2}$ weighting can provide bulk flow estimates with minimum predicted variance both by having very small side lobes and thus allowing one to simply determine the maximum radius for which a given survey can accurately determine the bulk flow.   

An additional advantage of the MV method is that it allows for constraints to be easily placed on the bulk flow moments using Lagrange multipliers.   Here we have imposed a constraint that ensures that the bulk flow moments are independent of the value of the Hubble constant $H_o$.  For the case where CF3 catalog velocities are calculated using $H_o=75$ km/s/Mpc, a value for which radial flows are roughly minimized, the constraint changes the bulk flow components by a few \kms and reduces the $\chi^2$ value from 10.49 to 9.59.  The probability of finding as large or larger a bulk flow is increased to 2.2\% from the 1.5\% obtained without the constraint.   We note that the current tension in the value of the Hubble constant makes it difficult to assess the magnitude of the radial flows in our local volume.  The effect of the constraint could be somewhat larger in a peculiar velocity catalog that has more significant radial flows or a more anisotropic distribution.

While the result of our analysis, that there is only a $\sim 2\%$ chance of obtaining the observed bulk given the parameters of the cosmological standard model, is similar to that of WFH, it is important to note the differences in both the data and the method used.   

First, the quantity of peculiar velocity measurements has increased dramatically; whereas WFH had $\sim$4,500 peculiar velocities of groups and individual galaxies in the COMPOSITE catalogue, the current analysis uses nearly 12,000, an increase of more than a factor of two.   The addition of new data hasn't significantly changed the direction of the bulk flow, but it has reduced it's magnitude;  the gaussian weighted bulk flow with $R=150$\hmpc\ (corresponding to $R=50$\hmpc\ in WFH) went from about $400$\kms\ in WFH to less than $300$\kms\ in the current analysis.   The addition of new data has also resolved an unexpected result from WFH;  they saw the bulk flow sharply \textit{increase} as the scale $R$ became large, a result that is difficult to provide a physical explanation for.   In Fig.~\ref{fig:bf} we see that in the current work the bulk flow decreases and then remains roughly constant with increasing $R$.  

Second, we have introduced a $r^{-2}$ weighting scheme  
which results in a window function with smaller side lobes than in the Gaussian weighting, indicating that bulk flows calculated with this weighting are less susceptible to velocity modes on scales smaller than the survey.    Given that they are only sensitive to scales as large or larger than $R$, they necessarily have smaller \textit{expectations} for the bulk flow, as seen in Fig.~\ref{fig:bf}.     However, we also see in Fig.~\ref{fig:bf} that the bulk flow for the $r^{-2}$ weighting is actually larger than that found using Gaussian weighting, so that the $\chi^2$ for the $r^{-2}$ weighting is significantly larger than for the Gaussian case.   In fact, the $\chi^2$ we calculate for the $r^{-2}$ weights are similar to those obtained by WFH for Gaussian weighting.   

It is important to note that disagreement with the standard cosmological model only occurs at the largest scales probed in this study, $\sim 150$\hmpc.   At smaller scales, the bulk flow magnitude is consistent with expectations.  Thus the question of whether the bulk flow is inconsistent with the standard cosmology depends strongly on precisely how the bulk flow is calculated.   It is not surprising, then, that there is disagreement on this question in the community, where different analyses can weigh information in different ways, even when using the same peculiar velocity catalogue.     Our result by itself is only suggestive of continuing tension with the standard model, but is not conclusive.   Ultimately, resolving the question of whether large scale flows are consistent with expectations will require additional measurements of peculiar velocities, particularly of objects with distances $\sim 150$\hmpc.   

\noindent{\bf Acknowledgements:} SP and RW have been supported in part by a grant from the Murdock Charitable Trust. We would like to thank Andrew Jaffe for his helpful comments.

\bibliographystyle{mn2e}
\bibliography{haf}

\begin{thebibliography}{34}
\expandafter\ifx\csname natexlab\endcsname\relax\def\natexlab#1{#1}\fi

\bibitem[{{Agarwal} {et~al.}(2012){Agarwal}, {Feldman}, \&
  {Watkins}}]{AgaFelWat12}
{Agarwal} S., {Feldman} H.~A., {Watkins} R., 2012, \mnras, 424, 2667

\bibitem[{{Bernal} {et~al.}(2016){Bernal}, {Verde}, \& {Riess}}]{BerVerRie16}
{Bernal} J.~L., {Verde} L., {Riess} A.~G., 2016, \jcap, 10, 019

\bibitem[{{Davis} {et~al.}(2011){Davis}, {Nusser}, {Masters}, {Springob},
  {Huchra}, \& {Lemson}}]{DavNusMas11}
{Davis} M., {Nusser} A., {Masters} K.~L., {Springob} C., {Huchra} J.~P.,
  {Lemson} G., 2011, \mnras, 413, 2906

\bibitem[{{Davis} \& {Scrimgeour}(2014)}]{DavScr14}
{Davis} T.~M., {Scrimgeour} M.~I., 2014, \mnras, 442, 1117

\bibitem[{{Dressler} {et~al.}(1987){Dressler}, {Faber}, {Burstein}, {Davies},
  {Lynden-Bell}, {Terlevich}, \& {Wegner}}]{DreFabBurDav87}
{Dressler} A., {Faber} S.~M., {Burstein} D., {Davies} R.~L., {Lynden-Bell} D.,
  {Terlevich} R.~J., {Wegner} G., 1987, \apjl, 313, L37

\bibitem[{Eisenstein \& Hu(1998)}]{EisHu98}
Eisenstein D.~J., Hu W., 1998, Astrophys. J., 496, 605

\bibitem[{{Feix} {et~al.}(2017){Feix}, {Branchini}, \& {Nusser}}]{FeiBraNus17}
{Feix} M., {Branchini} E., {Nusser} A., 2017, \mnras, 468, 1420

\bibitem[{{Feldman} \& {Watkins}(2008)}]{FelWat08}
{Feldman} H.~A., {Watkins} R., 2008, \mnras, 387, 825

\bibitem[{{Feldman} {et~al.}(2010){Feldman}, {Watkins}, \&
  {Hudson}}]{FelWatHud10}
{Feldman} H.~A., {Watkins} R., {Hudson} M.~J., 2010, \mnras, 407, 2328

\bibitem[{{Freedman}(2017)}]{Freedman17}
{Freedman} W.~L., 2017, Nature Astronomy, 1, 0121

\bibitem[{{Hellwing} {et~al.}(2018){Hellwing}, {Bilicki}, \&
  {Libeskind}}]{HelBilLib18}
{Hellwing} W.~A., {Bilicki} M., {Libeskind} N.~I., 2018, \prd, 97, 103519

\bibitem[{{Hellwing} {et~al.}(2017){Hellwing}, {Nusser}, {Feix}, \&
  {Bilicki}}]{HelNusFeiBil17}
{Hellwing} W.~A., {Nusser} A., {Feix} M., {Bilicki} M., 2017, \mnras, 467, 2787

\bibitem[{{Jaffe} \& {Kaiser}(1995)}]{JafKai95}
{Jaffe} A.~H., {Kaiser} N., 1995, \apj, 455, 26

\bibitem[{{Kaiser}(1988)}]{Kai88}
{Kaiser} N., 1988, \mnras, 231, 149

\bibitem[{{Lauer} \& {Postman}(1994)}]{LP94}
{Lauer} T.~R., {Postman} M., 1994, \apj, 425, 418

\bibitem[{{Macaulay} {et~al.}(2011){Macaulay}, {Feldman}, {Ferreira}, {Hudson},
  \& {Watkins}}]{MacFelFer11}
{Macaulay} E., {Feldman} H., {Ferreira} P.~G., {Hudson} M.~J., {Watkins} R.,
  2011, \mnras, 414, 621

\bibitem[{{Macaulay} {et~al.}(2012){Macaulay}, {Feldman}, {Ferreira}, {Jaffe},
  {Agarwal}, {Hudson}, \& {Watkins}}]{MacFelFerJaf12}
{Macaulay} E., {Feldman} H.~A., {Ferreira} P.~G., {Jaffe} A.~H., {Agarwal} S.,
  {Hudson} M.~J., {Watkins} R., 2012, \mnras, 425, 1709

\bibitem[{{Mukhanov} \& {Chibisov}(1981)}]{MukChi81}
{Mukhanov} V.~F., {Chibisov} G.~V., 1981, Soviet Journal of Experimental and
  Theoretical Physics Letters, 33, 532

\bibitem[{{Nusser}(2014)}]{Nusser14}
{Nusser} A., 2014, \apj, 795, 3

\bibitem[{{Nusser}(2016)}]{Nusser16}
---, 2016, \mnras, 455, 178

\bibitem[{{Nusser} \& {Davis}(2011)}]{NusDav11}
{Nusser} A., {Davis} M., 2011, \apj, 736, 93

\bibitem[{{Planck Collaboration} {et~al.}(2016{\natexlab{a}}){Planck
  Collaboration}, {Ade}, {Aghanim}, {Arnaud}, {Ashdown}, {Aumont},
  {Baccigalupi}, {Banday}, {Barreiro}, {Bartlett}, \& et~al.}]{PlanckCosPar16}
{Planck Collaboration}, {Ade} P.~A.~R., {Aghanim} N., {Arnaud} M., {Ashdown}
  M., {Aumont} J., {Baccigalupi} C., {Banday} A.~J., {Barreiro} R.~B.,
  {Bartlett} J.~G., et~al., 2016{\natexlab{a}}, \aap, 594, A13

\bibitem[{{Planck Collaboration} {et~al.}(2016{\natexlab{b}}){Planck
  Collaboration}, {Aghanim}, {Ashdown}, {Aumont}, {Baccigalupi}, {Ballardini},
  {Banday}, {Barreiro}, \& et~al.}]{PlanckXLVIII16}
{Planck Collaboration}, {Aghanim} N., {Ashdown} M., {Aumont} J., {Baccigalupi}
  C., {Ballardini} M., {Banday} A.~J., {Barreiro} R.~B.~., et~al.,
  2016{\natexlab{b}}, \aap, 596, A109

\bibitem[{{Riess} {et~al.}(2018){Riess}, {Casertano}, {Yuan}, {Macri},
  {Anderson}, {MacKenty}, {Bowers}, {Clubb}, {Filippenko}, {Jones}, \&
  {Tucker}}]{RieCasYuaMar18}
{Riess} A.~G., {Casertano} S., {Yuan} W., {Macri} L., {Anderson} J., {MacKenty}
  J.~W., {Bowers} J.~B., {Clubb} K.~I., {Filippenko} A.~V., {Jones} D.~O.,
  {Tucker} B.~E., 2018, \apj, 855, 136

\bibitem[{{Riess} {et~al.}(2016){Riess}, {Macri}, {Hoffmann}, {Scolnic},
  {Casertano}, {Filippenko}, {Tucker}, {Reid}, {Jones}, {Silverman},
  {Chornock}, {Challis}, {Yuan}, {Brown}, \& {Foley}}]{RieMarHofSco16}
{Riess} A.~G., {Macri} L.~M., {Hoffmann} S.~L., {Scolnic} D., {Casertano} S.,
  {Filippenko} A.~V., {Tucker} B.~E., {Reid} M.~J., {Jones} D.~O., {Silverman}
  J.~M., {Chornock} R., {Challis} P., {Yuan} W., {Brown} P.~J., {Foley} R.~J.,
  2016, \apj, 826, 56

\bibitem[{{Riess} {et~al.}(1995){Riess}, {Press}, \& {Kirshner}}]{RiePreKir95}
{Riess} A.~G., {Press} W.~H., {Kirshner} R.~P., 1995, \apjl, 445, L91

\bibitem[{{Rubin} {et~al.}(1976){Rubin}, {Thonnard}, {Ford}, \&
  {Roberts}}]{RubThoForRob76}
{Rubin} V.~C., {Thonnard} N., {Ford} Jr. W.~K., {Roberts} M.~S., 1976, \aj, 81,
  719

\bibitem[{{Scrimgeour} {et~al.}(2016){Scrimgeour}, {Davis}, {Blake},
  {Staveley-Smith}, {Magoulas}, {Springob}, {Beutler}, {Colless}, {Johnson},
  {Jones}, {Koda}, {Lucey}, {Ma}, {Mould}, \& {Poole}}]{ScrDavBlaSta15}
{Scrimgeour} M.~I., {Davis} T.~M., {Blake} C., {Staveley-Smith} L., {Magoulas}
  C., {Springob} C.~M., {Beutler} F., {Colless} M., {Johnson} A., {Jones}
  D.~H., {Koda} J., {Lucey} J.~R., {Ma} Y.-Z., {Mould} J., {Poole} G.~B., 2016,
  \mnras, 455, 386

\bibitem[{{Springob} {et~al.}(2014){Springob}, {Magoulas}, {Colless}, {Mould},
  {Erdo{\u g}du}, {Jones}, {Lucey}, {Campbell}, \& {Fluke}}]{SprMagCol14}
{Springob} C.~M., {Magoulas} C., {Colless} M., {Mould} J., {Erdo{\u g}du} P.,
  {Jones} D.~H., {Lucey} J.~R., {Campbell} L., {Fluke} C.~J., 2014, \mnras,
  445, 2677

\bibitem[{{Tully} {et~al.}(2013){Tully}, {Courtois}, {Dolphin}, {Fisher},
  {H{\'e}raudeau}, {Jacobs}, {Karachentsev}, {Makarov}, {Makarova},
  {Mitronova}, {Rizzi}, {Shaya}, {Sorce}, \& {Wu}}]{TulCouDol13}
{Tully} R.~B., {Courtois} H.~M., {Dolphin} A.~E., {Fisher} J.~R.,
  {H{\'e}raudeau} P., {Jacobs} B.~A., {Karachentsev} I.~D., {Makarov} D.,
  {Makarova} L., {Mitronova} S., {Rizzi} L., {Shaya} E.~J., {Sorce} J.~G., {Wu}
  P.-F., 2013, \aj, 146, 86

\bibitem[{{Tully} {et~al.}(2016){Tully}, {Courtois}, \& {Sorce}}]{CF3}
{Tully} R.~B., {Courtois} H.~M., {Sorce} J.~G., 2016, ArXiv e-prints

\bibitem[{{Watkins} \& {Feldman}(2015{\natexlab{a}})}]{WatFel15a}
{Watkins} R., {Feldman} H.~A., 2015{\natexlab{a}}, \mnras, 450, 1868

\bibitem[{{Watkins} \& {Feldman}(2015{\natexlab{b}})}]{WatFel15}
---, 2015{\natexlab{b}}, \mnras, 447, 132

\bibitem[{{Watkins} {et~al.}(2009){Watkins}, {Feldman}, \&
  {Hudson}}]{WatFelHud09}
{Watkins} R., {Feldman} H.~A., {Hudson} M.~J., 2009, \mnras, 392, 743

\end{thebibliography}

\end{document}